\documentclass{Interspeech2024}

\interspeechcameraready

\title{Fine-Tuning Automatic Speech Recognition for People with Parkinson's: An Effective Strategy for Enhancing Speech Technology Accessibility}

\name{Xiuwen}{Zheng}
\name{Bornali}{Phukon}
\name{Mark}{Hasegawa-Johnson}

\address{
  University of Illinois Urbana-Champaign\\
  Department of Electrical and Computer Engineering
  }
\email{xiuwenz2@illinois.edu, bornalip@illinois.edu, jhasegaw@illinois.edu}


\usepackage{graphicx}
\usepackage{array}
\usepackage{longtable}
\usepackage{subcaption} 

\begin{document}

\maketitle
\begin{abstract}
This paper enhances dysarthric and dysphonic speech recognition by fine-tuning pretrained automatic speech recognition (ASR) models on the 2023-10-05 data package of the Speech Accessibility Project (SAP), which contains the speech of 253 people with Parkinson's disease. 
Experiments tested methods that have been effective for Cerebral Palsy, including the use of speaker clustering and severity-dependent models, weighted fine-tuning, and multi-task learning.  Best results were obtained using a multi-task learning model, in which the ASR is trained to produce an estimate of the speaker's impairment severity as an auxiliary output.
The resulting word error rates are considerably improved relative to a baseline model fine-tuned using only Librispeech data, with word error rate improvements of 
37.62\% and 26.97\% compared to fine-tuning on 100h and 960h of LibriSpeech data, respectively. 

\end{abstract}

\section{Introduction}
Speech recognition technology has revolutionized how we interact with digital devices, and can be an especially useful computer input modality for people with motor disorders~\cite{friedoken1985voice,parker2006automatic}.  However, many individuals with motor disorders 
face challenges in effectively using these devices due to dysarthria and dysphonia, i.e., impaired articulation and phonation caused by neuromotor disorder~\cite{deller1991on,coleman1991computer}.

Despite ongoing efforts by researchers~\cite{asemi2019,caballeromorales2009,mustafa2014,sehgal2015model,shahamiri2014a,takashima2019,vachhani2018data,yu2018development}, current ASR systems encounter notable challenges in accurately recognizing dysarthric and dysphonic impaired speech primarily due to differences between impaired speech and typical speech and the unavailability of adequate training data. Initiatives like the Nemours database of dysarthric speech~\cite{menendez1996nemours}, the Universal Access speech corpus~\cite{kim2008dysarthric}, the TORGO database of acoustic and articulatory speech~\cite{rudzicz2012torgo}, the 
Neurovoz~\cite{moro2018study} and GITA~\cite{orozco2014new} Spanish-language corpora of speech with Parkinson's, 
and Project Euphonia~\cite{macdonald2021disordered} have been instrumental in advancing impaired speech recognition. However, while dysarthric ASR error rates
have dropped by a factor of three in the past fifteen years (\cite{geng2023use} vs.~\cite{sharma2010state}), non-dysarthric ASR error rates have dropped by a factor of five over the same period (\cite{hwang2023comparison} vs.~\cite{nguyen2005bbn}). The accelerated enhancements in non-dysarthric ASR can be partly credited to the greater accessibility of standardized data for both training and testing. To address these limitations, the Speech Accessibility Project (SAP~\cite{omalley2023uiuc}) is collecting, and distributing to researchers, a large corpus of impaired speech.  In this paper, we report the results of fine-tuning ASR using the first partial release of SAP data.

In the context of impaired speech recognition, fine-tuning a pre-trained model using transcribed impaired speech is a widely explored technique that has shown promising results for speech accessibility~\cite{green2021automatic}.  In particular, the wav2vec 2.0 base model has been fine-tuned to UA-Speech using low-rank adaptation~\cite{baskar2022speaker} and using system fusion methods in which wav2vec rescores a conformer or TDNN~\cite{hu2023exploring}.
Auxiliary speaker attributes can improve dysarthric ASR, e.g., hidden Markov model (HMM) ASR using UA-Speech was improved by first clustering speaker attribute vectors, then training cluster-dependent ASR for the speakers in each cluster~\cite{Christensen2014Auto}.  Deep neural network (DNN) ASR is improved by providing automatic dysarthria severity estimates as auxiliary inputs~\cite{liu2021recent} or outputs~\cite{geng2023use} of the ASR,
or by providing automatically inferred estimates of the visual~\cite{liu2020exploiting} or articulatory~\cite{liu2021recent} features corresponding to an acoustic dysarthric utterance.
Inspired by these results,
we aim to explore the potential benefits of fine-tuning the wav2vec 2.0 model by speaker clusters for dysarthric speech recognition.  We explore two types of speaker clusters-(1) Speaker feature-based. (2) Speech impairment severity-based. We have evaluated the performance of these experiments by comparing to a baseline wav2vec 2.0 ASR fine-tuned using only LibriSpeech data.

From the various experimental setups, it is observed that fine-tuning the wav2vec 2.0 base model on the SAP dataset yields a notable 36.70\% and 25.90\% improvement in Word Error Rate (WER) compared to fine-tuning on 100h and 960h of LibriSpeech data, respectively. 
Error rates for high-severity test speakers are harmed if the training corpus includes only high-severity speakers, but are improved if the training algorithm emphasizes speakers with a low or moderate degree of impairment. Optimum results are achieved using either cluster-weighted training (all training data are used, but some are weighted more than others) or a form of multi-task training in which speaker severity level serves as an auxiliary model output.  Cluster-weighted and multi-task training give identical low error rates averaged across the entire test set, and both give substantial improvements especially for the test speakers with the most severe impairment.





\section{Data}
\label{sec:dataset}

The Speech Accessibility Project (SAP) aims to support research and development in impaired speech recognition by collecting, transcribing, and sharing speech samples from contributors with disabilities.
The first partial data release was made available to members of the six participating institutions on 2023-10-05, and is scheduled for release to other researchers on 2024-04-05.  
We will refer to the SAP 2023-10-05 partial corpus as SAP-1005 throughout this paper.

SAP data sets differ in many ways from existing corpora of impaired speech.  First, unlike UA-Speech, Nemours, and TORGO, SAP is designed for speaker-independent ASR.  The 2023-10-05 data package includes 253 speakers, of whom 190 are allocated to a training set, 21 to a development set, and 42 to a test set. The development and test sets are each further divided into shared and unshared subsets, depending on whether or not the utterance text was also read by some of the training speakers; all results in this paper report WER using only the unshared subset of the test corpus.  Second, the types of impairment in SAP are quite different from those in existing corpora.  UA-Speech and TORGO include speakers with spastic or mixed dysarthria as symptoms of cerebral palsy (CP) or amyotrophic lateral sclerosis (ALS).  Although SAP will eventually include speakers with CP or ALS~\cite{omalley2023uiuc}, the 2023-10-05 data package includes only speakers with Parkinson's disease.  It has been demonstrated that even relatively mild Parkinson's disease can cause significant increases in the WER of an ASR~\cite{moro2019study}.  Some of these speakers exhibit hypokinetic dysarthria (reduced intelligibility caused by reduced articulatory movement): perceptual ratings by professional speech pathologists using the differential diagnostic scales of~\cite{darley1969differential} indicate that 63 of the 253 speakers have imprecise consonants, while 54 have less than normal intelligibility.  Dysphonia (disordered phonation) is a more frequent symptom: 207 of the 253 speakers have monotone pitch or loudness (monopitch or monoloudness), while 243 have speech that is rated as other than the normal level of overall naturalness.

In order to protect participant privacy, with limited exceptions, participant-provided metadata is not distributed with the corpus.  In particular, participant gender is not specified in the corpus distribution.

\section{Methods}
In this section, we present the methodologies adopted for fine-tuning the Wav2Vec 2.0 model on the SAP-1005 corpus, along with the speaker clustering strategies. We first introduce the architecture of the Wav2Vec 2.0 base model and then delve into the fine-tuning and speaker clustering strategies applied to enhance its performance in the context of dysarthric and dysphonic speech.

\subsection{Wav2Vec 2.0 pre-trained model} 
\label{sec:pretrain}

Wav2Vec 2.0~\cite{baevski2020wav2vec} is a set of Transformer-based speech recognition models
pre-trained using 
self-supervised learning.
The model initiates by processing the raw waveform $X$ of speech audio using a multilayer convolutional neural network (CNN), producing latent audio representations $z_t$ of 25ms duration each: $z_t = \text{CNN}(X)$. These representations are then simultaneously passed through a quantizer and a transformer. The quantizer selects a speech unit $q_t$ from a learned inventory of units: $q_t = \text{Quantizer}(z_t)$. Approximately half of the audio representations undergo masking before entering the transformer, which integrates information from the entire audio sequence, resulting in contextualized representations $c_t$: $c_t = \text{Transformer}(z_{\text{masked}})$. Finally, the output of the transformer is 
optimized contrastively to select
the appropriate quantized speech units corresponding to the masked positions.

During the fine-tuning phase, a randomly initialized new linear projection layer is added on top of the context network to project representation into \(C\) classes, which corresponds to the target vocabulary. The optimization process involves minimizing the Connectionist Temporal Classification (CTC) loss \(L_{CTC}\) ~\cite{graves2006connectionist} on labeled speech data. With a Transformer language model, fine-tuning the Wav2Vec 2.0 base model with the complete 960 hours of labeled LibriSpeech data yielded a 1.8\% WER on the LibriSpeech test-clean set. 


\subsection{Fine-tuning on impaired speech}

\begin{table}[t]
\centering
\scalebox{0.9}{
\begin{tabular}{llll}
\toprule
 \multicolumn{1}{l}{Split} & \multicolumn{1}{c}{All} &\multicolumn{1}{c}{Gender}& \multicolumn{1}{c}{Severity} \\
  & \multicolumn{1}{c}{\#spk$|$dur}& \multicolumn{1}{c}{\begin{tabular}[c]{@{}cc@{}} M & F\end{tabular}}& \multicolumn{1}{c}{\begin{tabular}[c]{cccc}VL & L& M & H\end{tabular}   }   \\ \midrule
  Train & \begin{tabular}[c]{@{}c@{}} 190$|$151.47h \end{tabular} & \begin{tabular}[c]{cc} 109 & 81\end{tabular} & \begin{tabular}[c]{cccc} 70 & 54 & 48 & 18\end{tabular}\\
  
  Dev & \begin{tabular}[c]{@{}c@{}} { }{ }21$|$23.32h\end{tabular} & \begin{tabular}[c]{cc} { }10 & { }11\end{tabular} & \begin{tabular}[c]{cccc} { }8 & { }{ }7 & { }{ }4 & { }{ }2\end{tabular}\\
  Test & \begin{tabular}[c]{@{}c@{}} { }{ }42$|$39.84h\end{tabular} & \begin{tabular}[c]{cc} { }26 & { }16\end{tabular} & \begin{tabular}[c]{cccc} 13 & 13 & 11 & { }5\end{tabular}\\
\bottomrule
\end{tabular}}
\captionsetup{margin=0.2cm}
\caption{Number of speakers in the Speech Accessibility Project Corpus, 2023-10-05 release.  Gender and impairment severity are automatically estimated using methods described in the text.}
\label{tab:table1}
\vspace{-3.0em}
\end{table}
To assess the effectiveness of the SAP-1005 corpus in dysarthric speech recognition, we first fine-tune an ASR to SAP data.  The wav2vec 2.0 base model, previously pre-trained using Librispeech 960h as described in Section~\ref{sec:pretrain}, is further fine-tuned using the training subset of SAP-1005.
Fine-tuning is performed using
the fairseq open-source repository\footnote{\url{https://github.com/facebookresearch/fairseq}}. 

\subsection{Speaker clustering}

Motivated by \cite{Christensen2014Auto}, where the authors observed noticeable enhancements in HMM-based dysathric ASR through selective speaker training based on speaker clusters, in this paper we investigate whether this observation extends to DNN-based speech recognition systems. 
We extract X-vectors \cite{Snyder2018xvectors}, one of the most widely used deep speaker embeddings, using the pre-trained VoxCeleb Xvector System 1a\footnote{\url{https://kaldi-asr.org/models/m7}} through kaldi. This system trained on augmented VoxCeleb 1 and VoxCeleb 2 corpus achieves a 3.1\% equal error-rate (EER) in the speaker recognition task and outperforms any other previous X-vector systems.
The wav2vec 2.0 base system described in Section~\ref{sec:pretrain} is fine-tuned separately for each speaker cluster in order to create speaker-cluster-dependent ASR. 
X-vectors are also extracted from the recordings of each speaker in the development and test set.   Each speaker in the development or test set is then assigned to the cluster-dependent ASR with the nearest X-vector centroid, which is used to transcribe their speech.

\subsection{Severity-dependent ASR}

Classical papers about hypokinetic dysarthric and dysphonia  (e.g.,~\cite{darley1969differential}) describe them using a relatively compact set of perceptual features (e.g., monopitch and monoloudness).
It is possible that speakers with similar levels of impairment severity may have acoustically similar speech, and that severity-dependent ASR models may therefore outperform severity-independent models.

Rather than depending on human severity annotations, we measure the impairment severity of a speaker according to the average character error rate (CER) with which the 
Wav2Vec 2.0 base system fine-tuned on Librispeech 960h transcribes that person's speech. Dividing the 190 training speakers into three equal-sized groups yields CER thresholds of approximately 10\% and 20\%. However, since there exists considerable inter-speaker variability within the most severe group, i.e., among those individuals with a CER exceeding 20\%, an additional threshold at 40\% is introduced. Consequently we use four severity levels: very low (VL: $<10$\% CER), low (L: $10-20$\% CER), median (M: $20-40$\% CER), and high (H: $>40$\% CER).
Table \ref{tab:table1} gives the frequency of each severity level. Subsequently, fine-tuning experiments are conducted on a severity basis, utilizing one or multiple clusters of severity.
Since the severity level of a development or test speaker cannot be determined without the availability of reference transcriptions, we perform experiments in which every category of dev and test speakers is transcribed using each of the four severity-dependent ASRs, in order to determine whether applying the wrong ASR to a test speaker increases or decreases the resulting WER.





\subsection{Fine-tuning with weighted clusters}

As an alternative to training an ASR using data from only a subset of speakers, we also train a set of models that are fine-tuned using the full training set, but with extra weight placed on the correctness of transcriptions for speakers of a given severity.
With $n$ denoting the total number of clusters, the general structure of the loss function is as follows:
\begin{equation}
L_{ctc,total} = \sum_{k=1}^{n} \omega_{k} \cdot L_{ctc,k} \
\end{equation}
where $\omega_k$ is a hyperparameter specifying the weight given to the $k^{\text{th}}$ cluster of training speakers, and $L_{ctc,k}$ is the corresponding CTC loss.

\subsection{Multi-task learning with severity classification}

Previous studies~\cite{geng2023use} have suggested that, instead of dividing the training data by severity level, dysarthric ASR can be improved by training the ASR using a multi-task training criterion in which the severity level of a speaker is estimated as an auxiliary output of the neural net.  
Employing the same multitask learning (MTL) strategy on the SAP-1005 corpus, we experiment with
an interpolation between the initial transcription loss, $L_{ctc}$, and the cross-entropy loss of an auxiliary network output estimating impairment severity, $L_{seve}$\footnote{Empirically, $\omega_{1} = \omega_{2} = \frac{1}{2}$}:
\begin{equation}
L_{MTL} = \omega_1\cdot L_{ctc}+ \omega_{2} \cdot L_{seve}\
\end{equation}
The auxiliary network is a linear transformation of the outputs of the last transformer layer, passed through a logistic sigmoid.  Three different placements of the auxiliary network were tested: ``first-token'' (linear transform of the first transformer output), ``max" (linear transform of the maximum, across time, of each element of the transformer output vector), and ``mean" (linear transform of the mean, across time, of each element of the transformer output).


\section{Results and Discussion}


\begin{table}[t]
\centering
\scalebox{0.9}{
\begin{tabular}{llll}
\toprule
 \multicolumn{1}{l}{Fine-tuning} &\multicolumn{1}{l}{Fine-tuning}& SA &
 \multicolumn{1}{c}{testWER} \\

  \multicolumn{1}{l}{Corpus}& \multicolumn{1}{l}{Split}&& \begin{tabular}[c]{@{ }ccc@{}}
 All & { }{ }Male &Female\end{tabular}      \\ \midrule
 LibriSpeech & 100h &N& \begin{tabular}[c]{@{}lll@{}} 42.53 & 47.96 & 34.85 \end{tabular}\\
 & 960h     &N& \begin{tabular}[c]{@{}lll@{}} \textbf{36.33} &\textbf{41.59} & \textbf{28.85} \end{tabular}      \\\midrule
 SAP-1005 & All&N   & \begin{tabular}[c]{@{}lll@{}} \textbf{26.92} & \textbf{30.08} & \textbf{22.45}\end{tabular}\\\cmidrule(lr){2-4}
 & Male      &N& \begin{tabular}[c]{@{}lll@{}} \textbf{29.62} & \textbf{31.95} & 26.36\end{tabular}\\
  & Female   &N& \begin{tabular}[c]{@{}lll@{}} 30.96 & 36.04 & \textbf{23.73}\end{tabular}\\\cmidrule(lr){2-4}
  & Male      &Y& \begin{tabular}[c]{@{}lll@{}} \textbf{29.88} & \textbf{32.00} & 26.89\end{tabular}\\
 & Female   &Y& \begin{tabular}[c]{@{}lll@{}} 30.96 & 35.91 & \textbf{23.97}\end{tabular}\\ 
\bottomrule
\end{tabular}}
\caption{WER (\%) of Wav2Vec 2.0 base models fine-tuned by different corpora and splits, using speaker clustering with two clusters.  The two clusters are described in the Table as ``Male'' and ``Female'' based on post-hoc perceptual evaluation.  SA=training data augmented using SpecAugment.}
\label{tab:table2}
\vspace{-1.5em}
\end{table}

The evaluation results, as presented in Table \ref{tab:table2}, demonstrate a remarkable improvement in Word Error Rate (WER) when fine-tuning on SAP-1005 compared to LibriSpeech. 
ASR fine-tuned using the SAP-1005 training set achieves 36.70\% less WER than ASR fine-tuned using Librispeech 100h, and 25.90\% less than ASR fine-tuned using LibriSpeech 960h. 
Unlike most previous published results for impaired ASR, this is not speaker-dependent training: These results demonstrate that fine-tuning using one set of speakers with Parkinson's (the SAP-1005 training set) substantially reduces WER for another set of speakers with Parkinson's (the SAP-1005 test set).


\begin{figure}[t] 
    \centering
    \begin{subfigure}{0.45\textwidth}
        \centering
        \includegraphics[width=\textwidth, trim=48 48 48 48, clip]{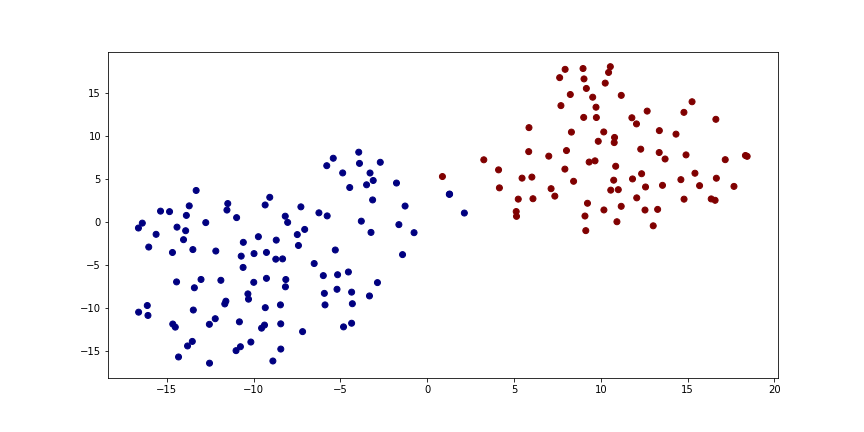}
        \label{fig:sub1}
    \end{subfigure}
    \begin{subfigure}{0.45\textwidth}
        \centering
        \includegraphics[width=\textwidth, trim=83.33 83.33 83.33 83.33, clip]{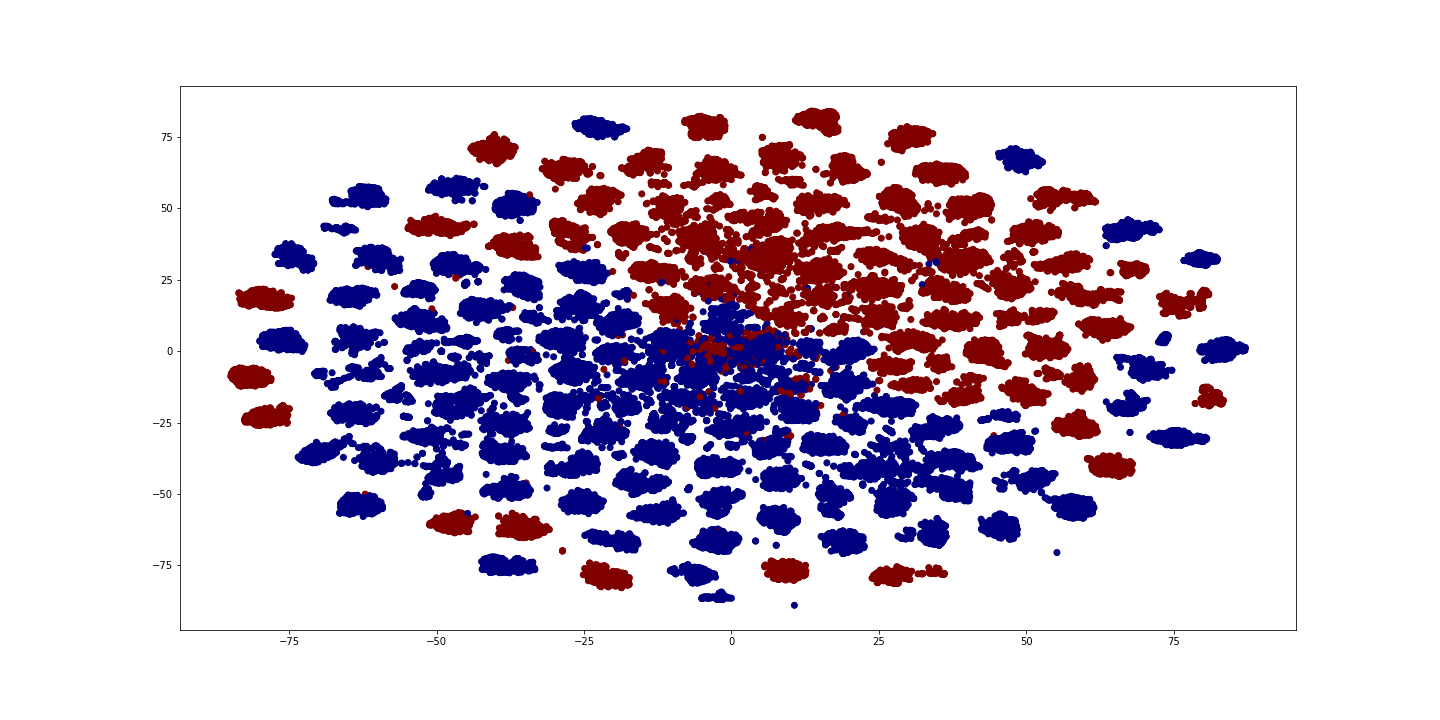}
        \label{fig:sub2}
    \end{subfigure}
    \caption{t-SNE plots of x-vectors with K-means clustering (K=2), using SAP-1005 training split. Top: speaker level; Bottom: utterance level.}
    \label{fig:overall}
    \vspace{-1.5em}
\end{figure}

X-vector clustering results in clusters of speakers each of whom sounds somewhat like the others in the same cluster.  When there are only two clusters, the two clusters seem to naturally divide the training speakers into two genders: post-hoc perceptual evaluation of the speakers in each cluster indicates that the speakers in one cluster are all male, while those in the other are all female, though this cannot be verified because the corpus distribution does not include gender metadata.
As shown in Figure \ref{fig:overall}, average x-vectors for each speaker are well separated between the clusters, and x-vectors computed from each utterance show a clear boundary between those from cluster 1 and those from cluster 2.
Table \ref{tab:table1} gives detailed number of speakers in each of the two clusters, where the clusters have been labeled ``Male'' and ``Female'' based on post-hoc perceptual evaluation. Based on this finding, we conduct fine-tuning experiments using two speaker clusters. In one experimental setting, SpecAugment~\cite{park2019specaugment} is used to increase the amount of training data in each cluster and to make the training data quantities similar in the two clusters. With data augmentation, approximately 151.47 hours of training data are selected for each experiment, aligning with all training data of the SAP-1005 corpus.

Matched testing (each test speaker is assigned to the cluster with the closest X-vector centroid) produces much lower WER than mismatched testing (in which each test speaker is assigned to the cluster with the farther X-vector centroid).
Specifically, fine-tuning on the male split leads to a 25.90\% WER reduction for males, but only 8.6\% for females, compared to the system fine-tuned on Librispeech. Conversely, fine-tuning on the female split yields a 17.75\% WER reduction for females, but 13.34\% for males. However, in contrast to the findings of Christensen et al.,~\cite{Christensen2014Auto}, where HMM-based speech systems trained by selecting speakers based on speech similarity outperformed those trained on all speakers, utilizing data from all speakers in our experiments led to enhanced model performance both generally and gender-specifically. 



The evaluation results of fine-tuning by severity are presented in Table \ref{tab:table3}.  Starting from a single cluster, we first compare models fine-tuned on the SAP-1005 ``VL," ``L" and ``M+H" subsets to ensure the amount of training data is comparable. Among these, while fine-tuning on ``VL" leads to the expected superior performance for ``VL" in the test split, fine-tuning on ``L" produces the best WER on every other severity levels except ``VL." 
Furthermore, we evaluate models fine-tuned on ``M" and ``H" splits separately. Compared to the ``M+H" model, the ``M" model benefits the overall performance, and the performance for test speakers with ``M" severity. On the other hand, test speakers with ``H'' severity are transcribed least accurately using an ASR fine-tuned using only ``H'' training speakers.  The set of ``H'' speakers in the training set contains 12.79 hours of speech, and previous publications suggest that 12.79 hours is sufficient to fine-tune an ASR with low WER~\cite{baevski2020wav2vec}, therefore we tentatively conclude that the very high WER of the system fine-tuned using ``H'' speakers is caused by their very high inter-speaker variability.

\begin{table}[t]
\centering
\vspace{-2em}
\scalebox{0.80}{
\begin{tabular}{lll}
\toprule
 \multicolumn{1}{l}{Fine-tuning} &\multicolumn{1}{l}{Fine-tuning}& \multicolumn{1}{c}{testWER} \\
  \multicolumn{1}{l}{Corpus}& \multicolumn{1}{l}{Split}& \multicolumn{1}{c}{\begin{tabular}[c]{lllll}
 All & { }{ }{ }{ }VL &{ }{ }{ }{ } L &{ } { }{ }{ }{ }M & { }{ }{ }{ }H\end{tabular}   }   \\ \midrule
 LibriSpeech & 100h & \begin{tabular}[c]{@{}lllll@{}} 42.53 & 23.72	&36.57	&61.49	&85.88\end{tabular}\\
 & 960h & \begin{tabular}[c]{@{}lllll@{}}  \textbf{36.33} & \textbf{18.36}	&\textbf{30.23}	&\textbf{54.16}	&\textbf{79.71}\end{tabular}\\\midrule
  SAP-1005 & All & \begin{tabular}[c]{@{}lllll@{}} \textbf{26.92} & 13.58	&21.15	&\textbf{41.10}	&\textbf{60.34}\end{tabular}\\
  & VL+L+M & \begin{tabular}[c]{@{}lllll@{}} \textbf{26.92} & \textbf{13.53}	&\textbf{20.79}	&41.58	&60.41\end{tabular}\\\cmidrule(lr){2-3}
  & VL+L & \begin{tabular}[c]{@{}lllll@{}} \textbf{27.51} & \textbf{13.64}	&\textbf{21.08}	&\textbf{42.57}	&62.93\end{tabular}\\
  & L+\underline{M+H} & \begin{tabular}[c]{@{}lllll@{}} 29.00 & 15.71	&23.48	&43.04	&\textbf{62.06}\end{tabular}\\
  & L+M & \begin{tabular}[c]{@{}lllll@{}} 28.75 & 15.30	&22.72	&43.32	&62.45\end{tabular}\\\cmidrule(lr){2-3}
  & VL & \begin{tabular}[c]{@{}lllll@{}} 30.97	&\textbf{14.46}	&24.15	&48.89	&70.68\end{tabular}\\
  & L & \begin{tabular}[c]{@{}lllll@{}} \textbf{30.13}	&15.83	&\textbf{23.52}	&\textbf{46.02}	&\textbf{65.50}\end{tabular}\\
  & \underline{M+H} & \begin{tabular}[c]{@{}lllll@{}} 33.90	&19.81	&28.98	&48.28	&66.99\end{tabular}\\
  & M & \begin{tabular}[c]{@{}lllll@{}} 32.92	&18.74	&26.90	&48.06	&68.00\end{tabular}\\
  & H & \begin{tabular}[c]{@{}lllll@{}} 51.70	&37.56	&46.51	&64.10	&82.39\end{tabular}\\
\bottomrule
\end{tabular}}
\caption{WER (\%) of Wav2Vec 2.0 base models fine-tuned by different corpora and splits, with severity clustering.}
\label{tab:table3}
\vspace{-3em}
\end{table}

\begin{figure}[t]
    \vspace{-2.5em}
    \centering
    \includegraphics[width=\linewidth]{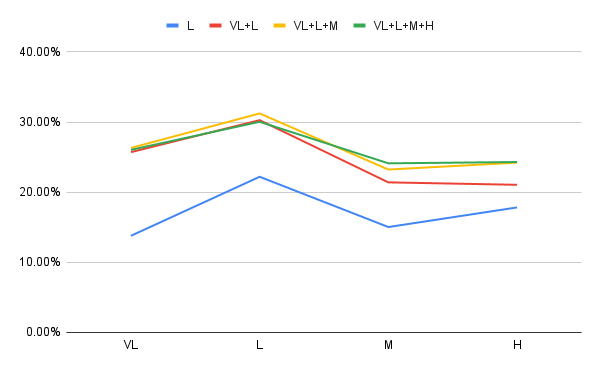}
    \caption{Relative improvements per severity level of the best performing models fine-tuned by 1, 2, 3 and 4 severity classes of  the SAP-1005 corpus, compared to the model fine-tuned by 960 hours of LibriSpeech.}
    \label{fig:wer_sev}
    \vspace{-1.5em}
\end{figure}

In general, feeding more data to the network leads to better performance, regardless of the severity level. Compared to the model fine-tuned on LibriSpeech 960h, Figure \ref{fig:wer_sev} shows the WER reduction of the best performing models fine-tuned with data from 1, 2, 3, or 4 distinct severity levels. While ``L" is the most powerful among all the models fine-tuned by a single cluster (30.13\% overall WER), as we enlarge the training data by combining two of the three major classes, the ``VL+L" model achieves the best overall WER of 27.51\% with an 8.70\% relative improvement compared to ``L" alone. Although adding ``VL" to ``L" leads to improvements among all severity levels, the influence decreases as the severity level increases, i.e., the WER reduction decreases from 11.93\%, 8.07\%, 6.37\%, to 3.22\% for ``VL," ``L," ``M," and ``H," respectively. This further indicates the mismatch between speech of different impairment severity levels. Combining three training groups (``VL+L+M'') further decreases the WER of individuals with ``M" and ``H" severity, without hurting performance for those with lower severity, leading to a 26.92\% overall test WER.  Adding ``H" on top of ``VL+L+M" does not further enhance overall performance:  there is no WER gain for test speakers in the ``H'' severity class, and WER changes are minimal for speakers in other categories.


\begin{table}[t]
\centering
\vspace{0.5em}
\scalebox{0.80}{
\begin{tabular}{lll}
\toprule
 \multicolumn{1}{l}{Fine-tuning} &\multicolumn{1}{l}{Fine-tuning}& \multicolumn{1}{c}{testWER} \\
  \multicolumn{1}{l}{Strategy}& \multicolumn{1}{l}{Parameters}& \multicolumn{1}{c}{\begin{tabular}[c]{lllll}
 All & { }{ }{ }{ }VL &{ }{ }{ }{ } L &{ } { }{ }{ }{ }M & { }{ }{ }{ }H\end{tabular}   }   \\ \midrule
  None & - & \begin{tabular}[c]{@{}lllll@{}} \textbf{26.92} & \textbf{13.58}&	\textbf{21.15}	&\textbf{41.10}	&\textbf{60.34}\end{tabular}\\\midrule
  Weighted & 1.2*L & \begin{tabular}[c]{@{}lllll@{}} \textbf{26.53}	&\textbf{13.28}	&\textbf{21.09}	&\textbf{40.45} &59.40\end{tabular}\\
  Fine-tuning & 1.6*L & \begin{tabular}[c]{@{}lllll@{}} 26.81	&13.74	&21.25	&40.78	&\textbf{59.16}\end{tabular}\\
  & 2.0*L & \begin{tabular}[c]{@{}lllll@{}} 27.20	&13.87	&21.84	&40.98	&60.49\end{tabular}\\
  & 3.0*L & \begin{tabular}[c]{@{}lllll@{}} 27.49	&14.32	&21.89	&41.42	&60.46\end{tabular}\\\midrule
Multitask  & first-token & \begin{tabular}[c]{@{}lllll@{}} \textbf{26.53}	&\textbf{13.36}	&\textbf{20.77}	&40.91	&\textbf{58.64}\end{tabular}\\
Learning  & max & \begin{tabular}[c]{@{}lllll@{}} 26.87	&13.53	&21.56	&\textbf{40.75}	&59.81\end{tabular}\\
  & mean & \begin{tabular}[c]{@{}lllll@{}} 27.10	&13.59	&21.37	&41.67	&60.10\end{tabular}\\
\bottomrule
\end{tabular}}
\caption{WER (\%) using SAP-1005 unshared test corpus. For weighted fine-tuning, the cluster along with adjusted weight are listed as parameters. For multitask learning, location of the auxiliary severity classifier is listed as a parameter.}
\label{tab:table4}
\vspace{-2.5em}
\end{table}

Table \ref{tab:table4} compares cluster-weighted fine-tuning and multi-task learning strategies to full-train-set fine-tuning.  By adding reasonable weights to the ``L" class, we achieve our best average WER of 26.53\% at $\omega_{L}=1.2$, while setting $\omega_{L}=1.6$ yields an improved 59.16\% WER for testing group ``H." Among all types of pooling layers for severity classification, first-token-based multitask learning yields the best overall WER (26.53\%), and the best WER for ``H''-severity test speakers (58.64\%). 




\section{Conclusion}
This work\footnote{This work utilizes resources supported by the National Science Foundation’s Major Research Instrumentation program, grant \#1725729, as well as the University of Illinois at Urbana-Champaign~\cite{HAL}} investigates the effectiveness of fine-tuning the wav2vec 2.0 model using the 2023-10-05 partial data release of the Speech Accessibility Project.  Our experiment demonstrates that fine-tuning using one group of speakers with Parkinson's disease (the SAP-1005 training speakers) substantially reduces WER for other speakers with Parkinson's disease (the SAP-1005 test speakers). Cluster-based fine-tuning does not further improve performance, but multi-task training, in which speaker severity level is an auxiliary output of the ASR, gives small additional benefits in performance, especially for test speakers with the most severe impairments.


\bibliographystyle{IEEEtran}
\bibliography{mybib}

\end{document}